\documentclass[useAMS,usenatbib,12]{article}
\usepackage[dvips]{graphicx}
\title {\textbf{ Interference detection in gaussian noise}}
\author {Raju R Baddi,$^1$\\ $^1$Raman Research Institute,C.V.Raman Avenue,Bangalore,560080,India\\ }
\date{}

\begin{document}

\label{firstpage}

\maketitle

\begin{abstract}

Interference detection in gaussian noise is proposed. 
It can be applied for easy detection and editing of interference 
lines in radio spectral line observations. One need not know the position of 
occurence or keep track of interference in the band. Results obtained on 
real data have been displayed.   

\end{abstract}


\section{Introduction}

Radio Frequency Interference(RFI) is a common problem during radio spectral 
line observations.
This RFI can be edited manually by inspecting the individual power spectra. 
RFI lines can appear at various positions in 
the spectrum depending upon the nature of the interference or the instrument settings. 
Here a few methods of detection have been proposed which apply to the 
power spectrum(i,e after the fourier transform of the recorded data) so that the RFI 
can be detected and edited using an algorithm.  Using such an 
RFI detection algorithm one can analyse huge amounts of
data with minimum time and effort.
The method has been described as it applies to radio spectral line observations 
using dual Dicke switching[1].

\section[]{The Radio spectral line observation}
In a typical radio spectral line observation one first decides upon the frequency at 
which the line will be observed. The bandwidth required to detect the 
line or lines and the period for which the observation will be made. 
This period decides the signal to noise ratio in the final spectrum.
The line is assumed to be observed with the procedure of dual Dicke switching. 
In this observation procedure  
the band is split into two equal parts. The spectral line is made to appear in these 
two parts alternately by appropriately tuning the telescope. This is usually done by
selecting two fixed local oscillator(LO) frequencies LO1 and LO2 over 
a period of time so that 
the gain characteristics of the instrument do not change significantly.  
Hence the observation consists of two types of 
spectra one with the spectral line appearing in the right part \& the other 
with the spectral line in the left part. The first is called the $T_{on}$ 
spectrum \& the other will be called the $T_{off}$ spectrum respectively.
Typical power spectra appear as shown in the fig-1. \\

\begin{figure}
\begin{center}
\includegraphics[width=120mm,height=80mm,angle=0]{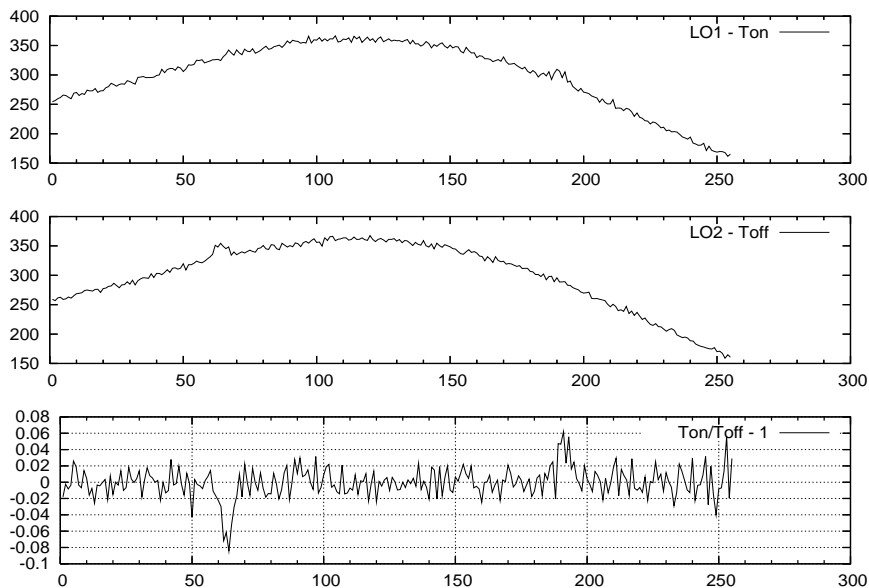}
\caption{Typical simulated power spectra with dual Dicke switching at tuning 
frequencies LO1 and LO2, and their combination $T_{on}/T_{off} - 1$. 
Abscissa is the channel number and the ordinate is arbitrary power. The spectral lines 
near channels 50(for Ton) and 200(for Toff) have been highly exagerated for the sake of clarity.}
\end{center}
\end{figure}
 
To detect the astronomical spectral line 
one is to first get rid of the background power(the band profile, refer fig-1, 
which is essentially 
the background radio power multiplied by the gain of the instrument) 
and normalize for the gain  
across the band. This can be acheived by simply doing a 
substraction of $T_{off}$ from $T_{on}$, since both have the same band profile. It is 
assumed that the bandshape of the instrument does not change appreciably for the 
two settings of the LO. The gain can be 
normalised by dividing the $T_{on}$-$T_{off}$ by $T_{off}$, the noisy features of 
$T_{off}$ in the denominator are not going to affect the ratio seriously 
since the derivative of $1/x$ goes as $1/x^{2}$(here $x$ is $T_{off}$). The main portion 
of the band where the lines are expected to appear has relatively higher 
magnitudes than the noise features. This can be seen in fig-1. 
The resulting output can be written as,

\begin{equation}
 PS=\frac{T_{on}-T_{off}}{T_{off}}
\end{equation} 

PS consists of two lines(bottom panel in fig-1), one from $T_{on}$ and the 
other from $T_{off}$. The spectral 
line from $T_{off}$ will be inverted since $-ve$ of $T_{off}$ was added to $T_{on}$. PS 
is assumed to be gaussian noise(or approximately gaussian) with an inherent 
baseline. Noise characteristics of real data have been displayed in fig-10. Since there is no 
difference between addition or substraction of gaussian noise 
due to its symmetry, both lead to gaussian noise again. 
After this PS is folded and averaged appropriately so that the inverted line overlaps with the 
one in the other part. Ofcourse a negative of the inverted part will be added.  
This completes the simple data processing. Many such spectra collected over time can be added 
to improve the signal to noise ratio which eventually results in spectral 
line detection. It is the property of gaussian noise that the rms(root mean 
square) of the averaged spectrum reduces by a factor of $1/\sqrt{n}$ when n spectra 
are averaged[1].

\subsection{Interference}
RFI is a common enemy of spectral line observation. It can be 
produced in a variety of ways, by electrical sparking, computers, 
electronic gadgets etc. The magnitude of interference can be both 
small as well as large. It is easy to detect the large interference 
while the small interference is the one which is to be tackled. 
A typical interference infected spectrum is as shown in the 
fig-2(some times it is low enough that it is distinguishable only in 
the ratio $(T_{on}-T_{off})/T_{off}$). \\

\begin{figure}
\begin{center}
\includegraphics[width=120mm,height=80mm,angle=0]{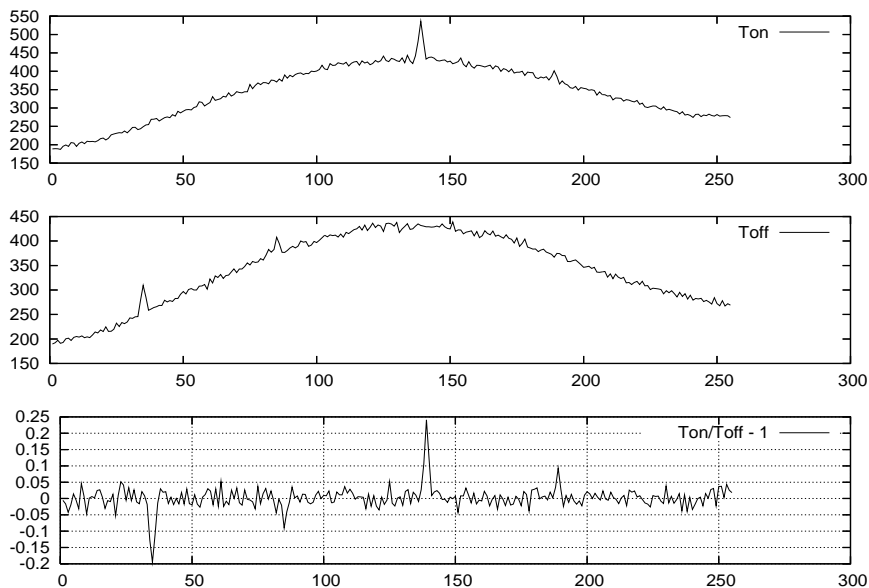}
\caption{Typical power spectra with a couple of interference lines appearing in Ton 
at $\sim$140 and $\sim$180. Here the amplitude 
has been chosen to be small, in reality it can be both much higher or smaller.}
\end{center}
\end{figure}

Different sources produce different types of RFI. For example 
the RFI generated by electrical sparking will produce a series 
of lines. In such cases the standard deviation of the data itself may 
increase drastically. Using this information such data can be discarded. 
While some other electronic equipment may produce  a single or multiple 
number of lines at some fixed position or 
positions in the band. In such cases one has to devise RFI excision 
techniques.

\section[]{RFI Detection}
The RFI lines appearing in the power spectrum can be recognized  
manually by carefully inspecting it. This would be laborious and time 
consuming. With the methods described here interference can be detected  
within the analysis program to auto recognize them. By an appropriate algorithm 
one can also have them edited(this is left to the user). This reduces manual 
labor as well as results in speedy data anaysis. In the methods proposed 
here the following assumptions have been made, 

\begin{enumerate}

\item \ The RFI can be observed in 
a short integration time where as the detection of the astronomical signal 
would require a much longer integration time. In other words the spectrum 
subjected to interference detection satisfies the criteria that 
the astronomical line is smaller than or comparable to the standard deviation 
of the spectra.

\item \ The RFI is narrow band, i.e only a small portion or portions of the band 
are infected with it. It is necessary to use a portion of the band to find the 
standard deviation of the spectrum. 
This parameter is necessary in the detection of interference. The standard deviation 
is an important quantity relevant to noise. 

\item \ It is also not necessary that the RFI be time variable. If the spectrum has 
interference as per the technique used here the associated channels will be flagged 
or else they will pass unflagged. 

\end{enumerate}

The technique to detect RFI depends on the properties of gaussian noise. When pure gaussian 
noise across N channels is considered there exists an expected maximum value that can occur 
across the channels. Similarly there exists for N channels a maximum for the difference 
between two adjacent channels. Lastly another quantity of interest is the added difference 
which is the sum of consecutive differences of similar sign( i.e the adjacent differences could 
be either -ve or +ve ). Suppose these difference signs go 
like + - + - + + - , then the fifth and sixth differences will be added to form one quantity. 
These quantities can be calculated theoretically and 
used to inspect the gaussian noise. Typical interference being sharp and narrow would produce, 
large channel values or differences between adjacent channel values that deviate considerably 
from the expected values. On such an occurence the corresponding channel or channels can be 
flagged for interference. The first quantity, expected maximum can be of use when the spectrum 
has reasonably zero mean all across the spectrum, in other words it has zero baseline,  
so is of less importance. The other two 
quantities are free from this defect as they are differences between adjacent channels which 
more or less have the same DC value. The methods using these three quantities have been named as 
direct,difference and added difference methods. The added difference traps 
those interference lines that escape the difference method due to their being slightly broad. Plots of 
these quantities signifying there use to detect interference have been given in the following 
sections.   

\subsection{Direct Method}
This is the simplest of all the methods.
In this method the spectrum of $N_{data}$ channels is divided into $n_{blk}$ 
number of blocks. Each block has now 

\begin{equation}
 N_{blk}=\frac{N_{data}}{n_{blk}}
\end{equation}

number of data points. The block with the minimum standard deviation 
will be considered to be 
the representative of good data, meaning to say free of any interference.
This is consistent in the sense that in the regions without any signal 
one has only noise, any interference introduced into this data will increase  
the standard deviation through increase in the mean value. The minimum standard 
deviation block will be used as a reference to calculate the required parameters to 
detect interference as well as to edit and repair it(one can replace the edited 
portion appropriately with noise of same standard deviation as that in the reference block). 
The absolute maximum value in this block can be used to generate a limit on 
the maximum absolute value that could be encountered in the other blocks by 
choosing a suitable multiplicative factor. This can also be done 
as the expected maximum is directly proportional to the standard deviation. Or 
else use the formula given below for $X_{mx}$ along with the standard deviation in 
the reference block and the total number of channels($N_{data}$). 
It should be noted that a constant muliplicative factor($>1$, say 1.1 - 1.3) 
has to be 
used with the expected maximum value to allow for the realistic departures from the 
expected value.  If any block has any channel or channels with absolute value 
greater than this, then those channels in that block can be flagged for interference. 
The expected maximum absolute value for a N channel gaussian noise with zero mean  
is given by

\begin{equation}
\left<X_{mx}\right> = \sigma \sqrt{2}\ \int_0^1 erf^{-1} (s^{\frac{1}{N}})\ ds 
\end{equation}
which has a convenient approximation as,
\begin{equation}
\left<X_{mx}\right>\approx \sigma \sqrt{2}\ erf^{-1}(1 - \frac{0.5833}{N})
\end{equation}
 
Derivation of this has been discussed in the appendix.
This method requires that the spectrum under investigation be symmetrically 
distributed about the zero line. This can be 
brought about by calculating the mean of all the channels of the data and 
substracting it from each channel value of the spectrum. Also in some cases 
a baseline should be removed from the 
spectrum before subjecting to RFI detection. In the case where the 
interference is present this 
is difficult since the fitted curve will be baised by the interference line.  
If one is sure that the spectrum has no baseline then one can use this 
method or else omit it. However the following methods are more tolerant towards 
this. 

\subsection{Difference Method}
The difference method is a simple method of interference detection. 
In this method adjacent channel differences all across a single spectrum 
will be found. For a N channel gaussian noise there exists an upper limit 
on the maximum expected difference. Which will be used as a reference to 
limit the maximum value that could be encountered across the N channels. 
Further the spectrum is divided into $n_{blk}$ blocks.  
Again the reference block will be the minimum standard deviation block. 
In each block the adjacent differences are found. A typical plot of 
differences of an interference infected $T_{on}/T_{off} - 1$ spectrum is 
shown in the fig-3 which relates to the spectrum in fig-2.
\begin{figure}
\begin{center}
\includegraphics[width=120mm,height=80mm,angle=0]{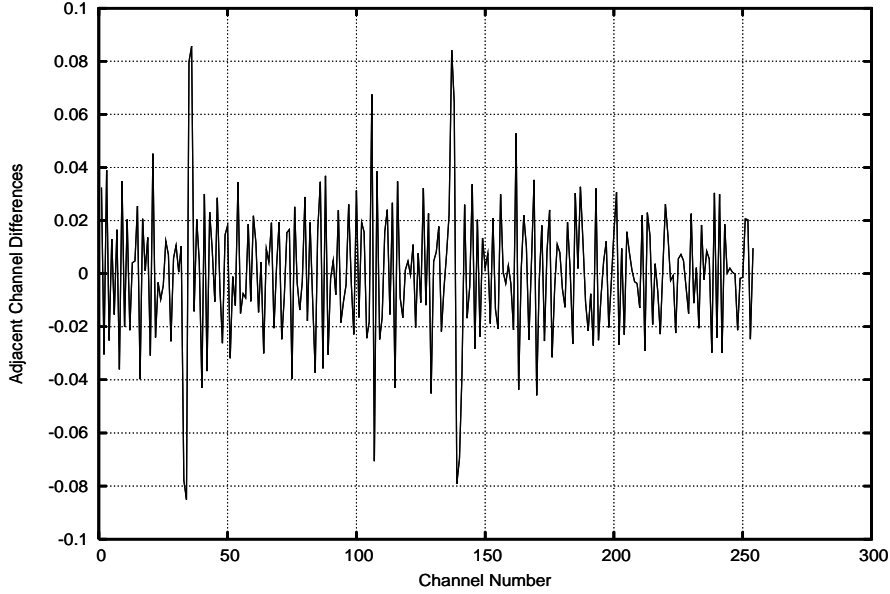}
\caption{A plot of differences across the spectrum for a typical interference
infected $(T_{on}-T_{off})/T_{off}$. The large differences in 
the interference infected region can be noticed.}
\end{center}
\end{figure}
Obviously the differences in the interference infected channels are much higher 
than the differences in the other channels. The expected value of the maximum 
difference($<D_{mx}>$) for gaussian noise with N data points is given by the relation,

\begin{equation}
\left<D_{mx}\right> \approx 2 \sigma\!\!\int_{0}^{1}\!\!erf^{-1}(s^{\frac{1}{N}})\>{{d}}s = \sqrt{2} \left<X_{mx}\right>
\end{equation} 

where $\sigma$ can be taken from reference block.
The derivation of this has been discussed in the appendix.\\

\subsection{Added Difference Method}
The added difference method is similar to the difference method 
except now the consecutive differences with same sign will be 
added to form a new array of quantities.  
Suppose the difference signs go like + - + - + + - , then the 
fifth and sixth differences will be added to form one quantity. 
The plot of such an array for a typical interference infected 
spectrum is shown in the fig-4 relating to fig-2. The improvement of interference detection criteria is clearly 
visible in this plot. Added difference method traps those 
interference lines which escape the difference method due to 
their being slightly broad.  For this method the 
constraining limit($<Q_{mx}>$) can be obtained by finding the maximum added 
difference in the reference block and use a suitable multiple($>1$) of this 
or else use the empirical formula below,

\begin{equation}
<Q_{mx}> = 2\ \sigma \ erf^{-1} \left( 1 - \frac{0.5833}{2 N} \right) .
\end{equation}  \\

\begin{figure}
\begin{center}
\includegraphics[width=120mm,height=80mm,angle=0]{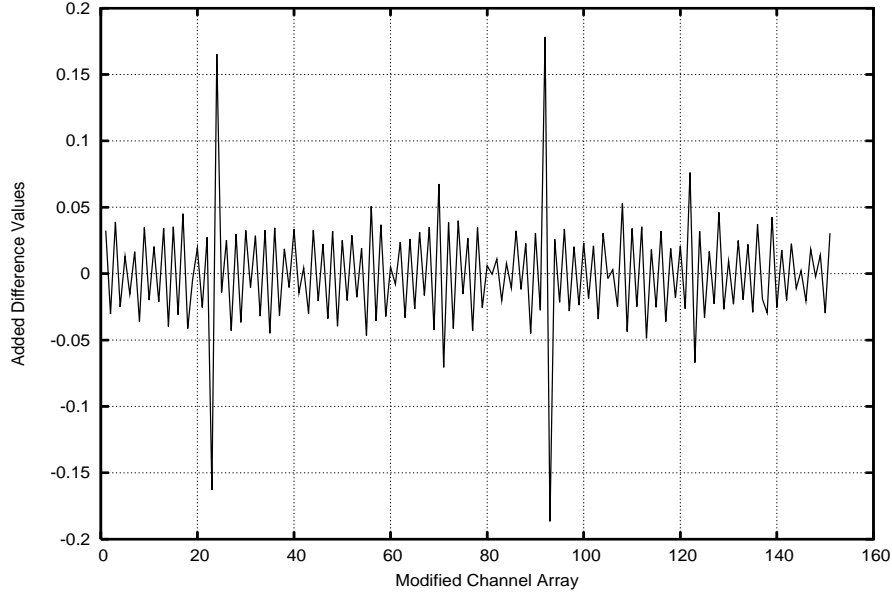}
\caption{A plot of added-differences array for the previous interference infected
 $(T_{on}-T_{off})/T_{off}$  spectrum. The improvement in the detection 
criteria compared to the differences in fig-3 is clearly apparent.}
\end{center}
\end{figure}

\appendix
\section[]{Derivation of the expected maximum value for the gaussian distribution}

\begin{figure}
\begin{center}
\includegraphics[width=120mm,height=28mm,angle=0]{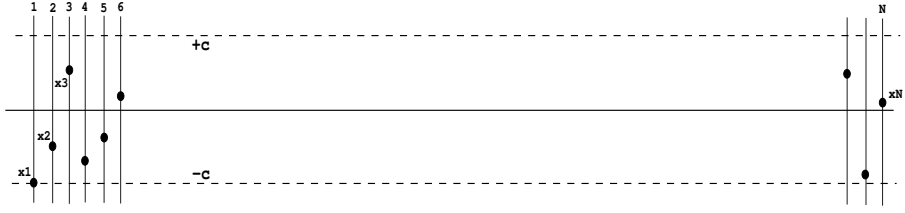}
\caption{Calculating the maximum expected value in N channel gaussian noise.}
\end{center}
\end{figure}

Considering N channels which produce gaussian noise(fig-5) of the same standard deviation 
$\sigma$. The probability that an absolute maximum value of c is produced in these 
channels is 

\begin{equation}
p(c) = \frac{2^N N e^{- \frac{c^2}{{2 \sigma^{2}}}}}{{\left( \sigma \sqrt{2 \pi} \right)}^{N}} \!\!\int_{0}^{c} e^{- \frac{{x_2}^2}{{2 \sigma^{2}}}}{{d}}x_{2} \!\!\int_{0}^{c} e^{- \frac{{x_3}^2}{{2 \sigma^{2}}}}{{d}}x_{3} \dots {{d}}x_{N}  
\end{equation} 
	   
which is nothing but the integral over all the possibilities over the remaining N-1 channels 
keeping one of the channels fixed at c. The factor $2^{N}$ is due to the fact that both +ve 
\& -ve values are possible in each channel. The remaining N-1 channels are allowed to take all 
the values between -c and +c. N in the numerator appears as there are N-ways to fix a 
channel value to c. Each of the above integrals evaluates to $ \sigma \sqrt{\pi/2} erf(c/{\sigma \sqrt{2}}) $ resulting 
in,  

\begin{equation}
p(c) = \frac{2 N e^{- \frac{c^2}{{2 \sigma^{2}}}}}{\sigma \sqrt{2 \pi}}{\left[ erf {\left(\frac{c}{\sigma \sqrt{2}} \right)} \right]}^{N-1} 
\end{equation}

Now the maximum expected value can be written as,
 
\begin{equation}
\left<X_{mx}\right> = \!\!\int_{0}^{\infty} c\ p(c) {{d}}c 
\end{equation}

\begin{equation}
\left<X_{mx}\right> = \frac{2 N}{\sigma \sqrt{2 \pi}}\!\!\int_{0}^{\infty} c\ e^{- \frac{c^2}{2 \sigma^{2}}}{\left[ erf {\left(\frac{c}{\sigma \sqrt{2}} \right)} \right]}^{N-1}{{d}}c 
\end{equation}

which can also be written using the derivative of $ S =  \left[erf\left(\frac{c}{\sigma \sqrt{2}}\right)\right]^{N} $ and expressing c in terms of S as

\begin{equation}
\left<X_{mx}\right> = \sigma \sqrt{2}\!\!\int_{0}^{1} erf^{-1}\left(S^{\frac{1}{N}} \right){{d}}S 
\end{equation}

which has an approximation
\begin{equation}
\left<X_{mx}\right> \approx \sigma \sqrt{2}\ erf^{-1} \left(1 - \frac{0.5833}{N} \right)  
\end{equation}

which has been derived by considering the series expansion for $erf^{-1}(S)$ . 

\begin{equation}
\int_{0}^{1} erf^{-1}\left( S^{1/N} \right){{d}}S = \sum_{k=0}^{\infty} \frac{C_{k}}{2k + 1}\left(\frac{\sqrt{\pi}}{2}\right)^{2k+1}\int_0^{1} S^{\frac{2k+1}{N}}{{d}}S
\end{equation}

\begin{equation}
\begin{array}{lcl}
\int_{0}^{1} erf^{-1}\left( S^{1/N} \right){{d}}S &=& \sum_{k=0}^{\infty} \frac{C_{k}}{2k + 1}\left(\frac{\sqrt{\pi}}{2}\right)^{2k+1} \left[\frac{S^{\frac{2k+N+1}{N}}}{{\frac{2k+N+1}{N}}}\right]_{0}^{1}\ \\ \\ 
&=&  \sum_{k=0}^{\infty} \frac{C_{k}}{2k + 1}\left(\frac{\sqrt{\pi}}{2}\right)^{2k+1} \frac{N}{2k+N+1}
\end{array}
\end{equation}

Considering $erf^{-1}(z)$, such that,

\begin{equation}
z^{2k+1} = \frac{N}{2k+N+1}
\end{equation}

\begin{equation}
z = \left[ 1 + \frac{2k+1}{N} \right]^{\frac{-1}{2k+1}}
\end{equation}

Since N is normally large and the higher(k) order terms in the series contribute less and less. 
The above equation has a simple approximation,

\begin{equation}
z = \left(1 - \frac{1}{N} \right)
\end{equation}

A better approximation is obtained by binomially expanding(16) and considering the first 5 terms. 
This introduces +ve errors in the lower order terms but helps to account for the truncated higher order terms. 
In this approximation 2k+1 $\sim$ N for all the terms. This yields the approximation,

\begin{equation}
z = \left(1 - \frac{0.5833}{N}\right)
\end{equation}\\

\section[]{Derivation of the expected maximum-difference value for the gaussian distribution}

To get the maximum expected difference between two adjacent channels in a N channel gaussian 
noise spectrum, first the probability of occurence of a difference of d in two channels 
is calculated(refer fig-6). Further it is assumed that each of the adjacent channels behave approximately 
independently and the two channel probability distribution is applied to the N channels. This 
approximation becomes better when one considers a large number of channels as can be seen 
by the analysis of a 3 channel gaussian noise.

\begin{figure}
\begin{center}
\includegraphics[width=50mm,height=40mm,angle=0]{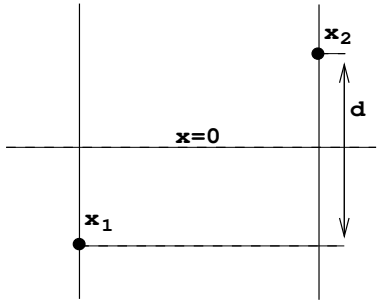}
\caption{Calculating the probability of occurence of a difference of d in 2 channels with N=2. }
\end{center}
\end{figure}

\begin{equation}
p_2(d) = \frac{1}{2 \pi \sigma^2} \!\!\int_{-\infty}^{+\infty} \ e^{-\frac{x^2 + (d+x)^2}{2 \sigma^2}}{{d}}x
\end{equation} 

so the probability density for a two channel difference is

\begin{equation}
p_2(d) = \frac{1}{2 \sigma \sqrt{\pi}} e^{-\frac{d^2}{4 \sigma^2}}
\end{equation}

Now the 3 channel probability analysis is as follows,

\begin{figure}
\begin{center}
\includegraphics[width=50mm,height=40mm,angle=0]{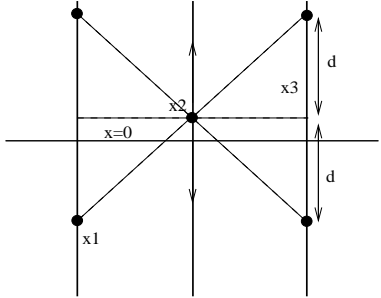}
\caption{Calculating the probability of occurence of a difference of d in two channels with N=3.}
\end{center}
\end{figure}

\begin{figure}
\begin{center}
\includegraphics[width=80mm,height=125mm,angle=-90]{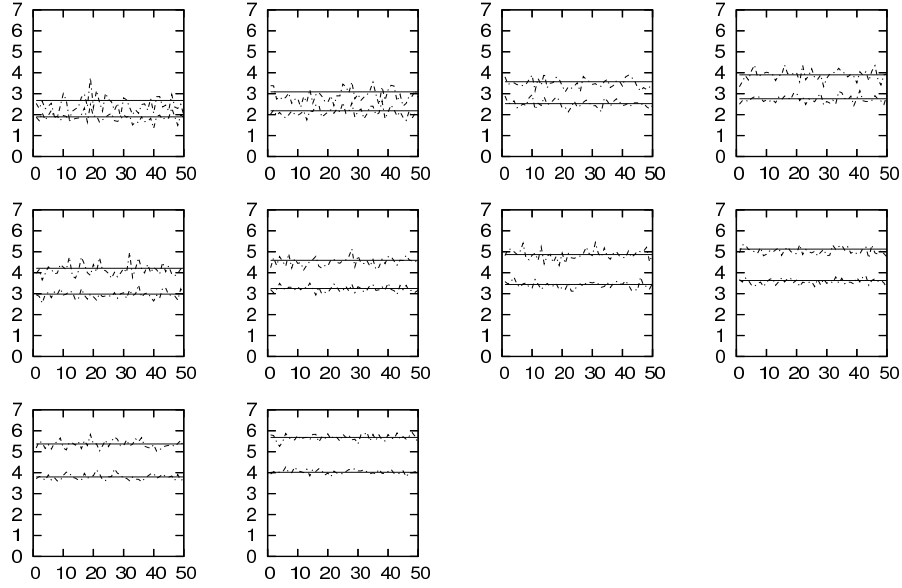}
\caption{Comparing the expected maxima and difference maxima for $\sigma=1$, from top left corner N=10,20,50,100,200,500,1000,2000,4000,10000. The lower comparision is of $X_{mx}$ and the upper comparision is of $D_{mx}$. The straight line is that calculated theoretically using equations given here (12) \& (28), while those obtained from octave('randn' function) are the broken lines. Each value is an average of 5 maxima samples. The abscissa is the trial number.}
\end{center}
\end{figure}

In the fig-7, x1 \& x2 are fixed such that the difference between them is d. x3 is 
allowed to take any value such that the its magnitude of difference with x2 does not 
exceed d. Next the same is repeated to account the other possibility of x2 and x3 
being fixed and x1 being free to take variable values. The net probability would be 
integrating over x2 from $-\infty$ to $+\infty$ . So we can write this net probability 
density as,

\begin{equation}
p_{3_2}(d) = \frac{2 \cdot 2}{(\sigma \sqrt{2 \pi})^3}\!\!\int_{-\infty}^{+\infty}\ e^{-\frac{x^2 + (x+d)^2}{2 \sigma^2}}\!\!\int_{-d}^{+d}\ e^{\frac{-(x+y)^2}{2 \sigma^2}}{{d}}y \ \ {{d}}x 
\end{equation}

\begin{equation}
p_{3_2}(d) = \frac{1}{\pi \sigma^2}\!\!\int_{-\infty}^{+\infty}\ e^{-\frac{x^2 + (x+d)^2}{2 \sigma^2}}\left( erf\left(\frac{d+x}{\sigma \sqrt{2}}\right) + erf\left(\frac{d-x}{\sigma \sqrt{2}}\right) \right)  \ \ {{d}}x 
\end{equation}

The factor of 2 in the numerator is due to the fact that d in the first integrand takes 2 values 
+ve and -ve and in both the cases it integrates to the same amount. Another 2 is due to the interchange 
of roles between x1 and x3. Now when a similar analysis 
is done by considering 4 channels and taking the difference in two channels independently, we 
see that the probabilities are approximately same for occurence of a difference d in the two 
cases. The probability in the case of 4 channels, but pair wise is

\begin{equation}
p_{4_2}(d) = \frac{2 \cdot 2}{(\sigma \sqrt{2 \pi})^2}\!\!\int_{-\infty}^{+\infty}\ e^{\frac{-x^2 + (x+d)^2}{2 \sigma^2}} \!\!\int_{-d}^{+d} p_2(y)\ {{d}}y \ {{d}}x
\end{equation}

\begin{equation}
p_{4_2}(d) = \frac{2}{ \pi\sigma^2}\!\!\int_{-\infty}^{+\infty}\ e^{\frac{-x^2 + (x+d)^2}{2 \sigma^2}} erf \left(\frac{d}{2 \sigma}\right) \ {{d}}x = \frac{2}{\sigma \sqrt{\pi}} \ e^{-\frac{d^2}{4 \sigma^2}}\ erf\left(\frac{d}{2 \sigma}\right)
\end{equation}

This calculation can also be done by applying $p_{2}(d)$ to both pairs. 
It can be checked(here checked numerically) that for a given value of d,

\begin{equation}
p_{3_2} \approx p_{4_2}
\end{equation}

Hence one can assume that the channels behave approximately independently pair wise in a 
series of N channels. By considering the difference probability distribution for 2 channels and 
treating the N-1 channel pairs as new single channels in which the occurence value is not 'x' 
but 'd' we can write using the same derivation as for $X_{mx}$ for the maximum expected difference 
value in a N channel gaussian noise as,

\begin{equation}
\left<D_{mx}\right> = \frac{N-1}{\sigma \sqrt{\pi}} \!\!\int_{0}^{\infty}\ d\ e^{-\frac{d^2}{4 \sigma^2}}\left[ erf\left(\frac{d}{2 \sigma}\right)\right]^{N-2} {{d}}d
\end{equation}

which again using $S = \left[ erf\left(\frac{d}{2 \sigma}\right)\right]^{N-1} $ can be 
written as,

\begin{equation}
\left<D_{mx}\right> = 2 \sigma \!\!\int_{0}^{1} erf^{-1}\left(S^{\frac{1}{N-1}} \right){{d}}S 
\end{equation}
 
For large N the -1 in the exponent can be dropped and simply one can use from (11)

\begin{equation}
\left<D_{mx}\right> = \sqrt{2} \left<X_{mx}\right>
\end{equation}\\

The derivations given here are of the author. He is unaware of any such derivations or results.\\

\begin{figure}[here]
\begin{center}
\includegraphics[width=110mm,height=85mm,angle=0]{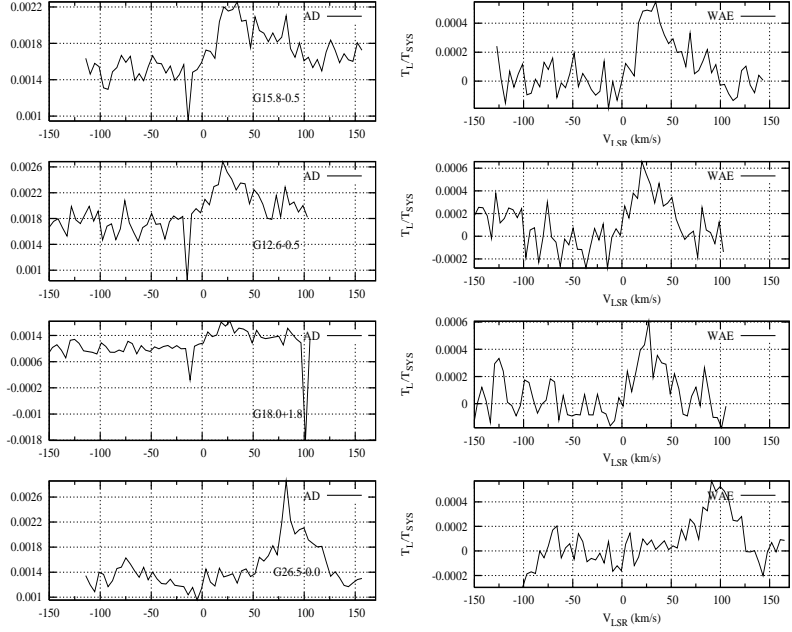}
\caption{The results obtained using difference and added difference methods on spectral line 
observations at ~328 MHz using Ooty Radio Telescope(ORT) towards 4 Galactic positions, inset left. 
WAE -collapse with automated editing and baseline removal, 
AD - collapse of actual data. The right 
plots are the auto-edited versions of the left plots. Interference detection \& editing 
was performed on 10-sec data. The collapse on the right consists of hundreds of such spectra. 
During editing the interference infected channels 
were replaced by noise of the same standard deviation as in the reference block + a straight line 
connecting adjacent channels on the two sides of the interference line. }
\end{center}
\end{figure}

\begin{figure}[here]
\begin{center}
\includegraphics[width=60mm,height=48mm,angle=0]{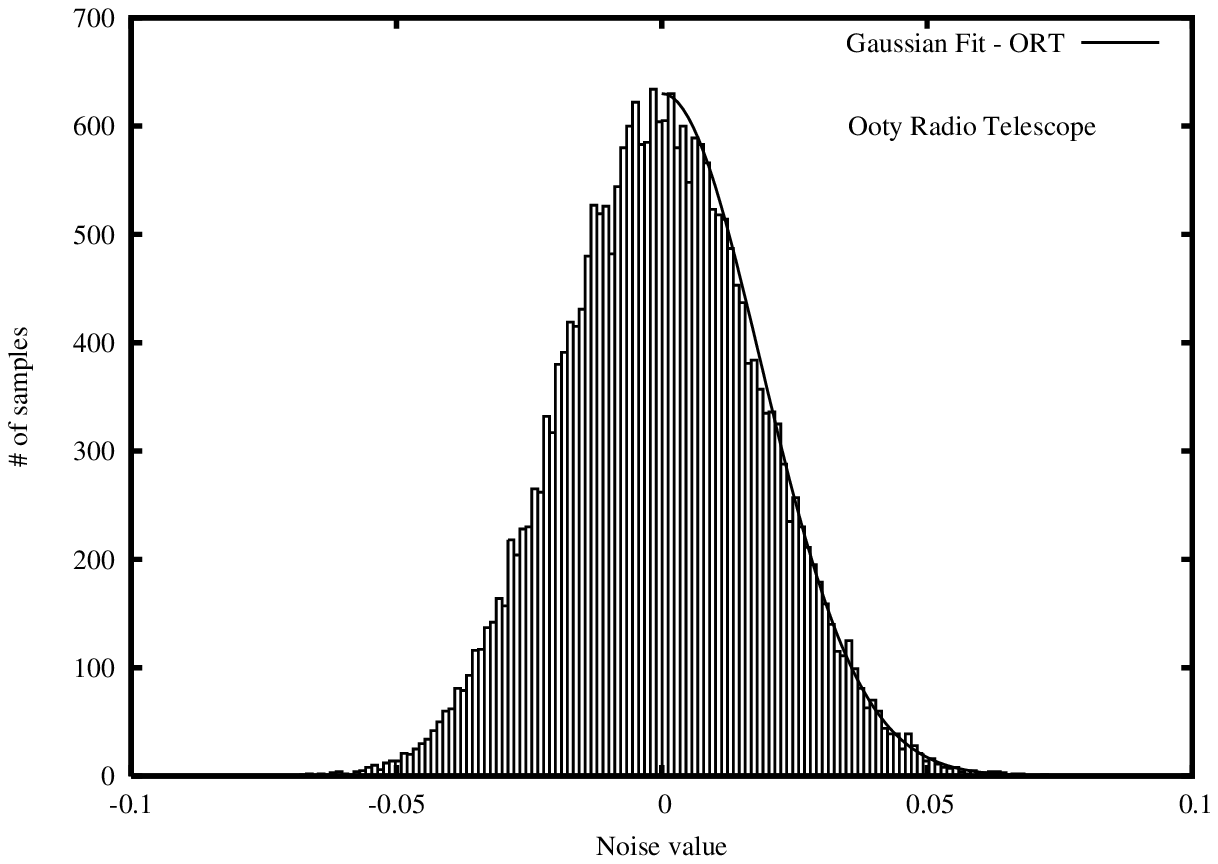}
\includegraphics[width=60mm,height=48mm,angle=0]{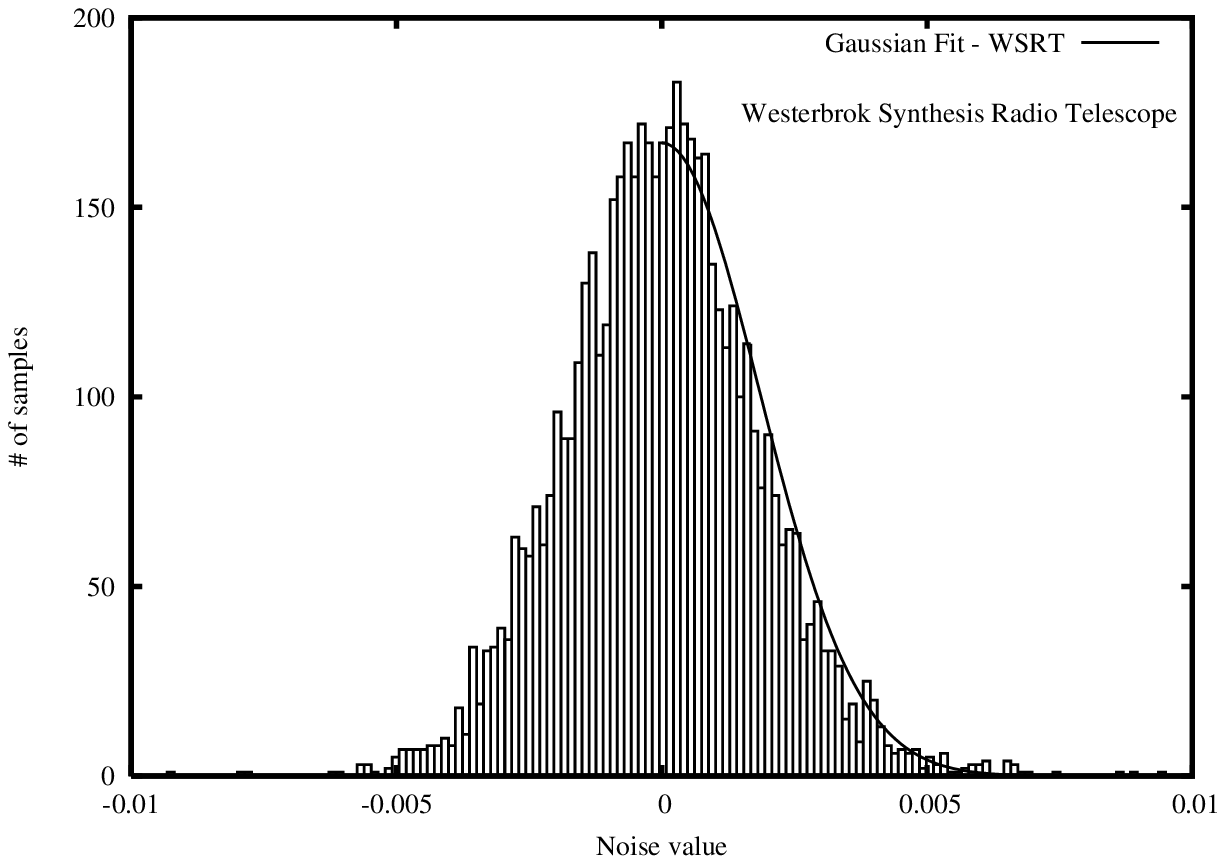}
\caption{Noise sampled from a large collection(ORT-100, WSRT-23) of Ton/Toff - 1 after baseline 
removal(refer fig-1). For ORT Ton and Toff correspond to 1-second spectrum, where as for WSRT its 1-minute.}
\end{center}
\end{figure}

\label{lastpage}

\end{document}